# Two-dimensional nonlinear modes and frequency combs in bottle microresonators

Y. V. KARTASHOV[1,2,*], M. L. GORODETSKY[3,4], A. KUDLINSKI[5], AND D. V. SKRYABIN[2,3,6]

[1]*Institute of Spectroscopy, Russian Academy of Sciences, Troitsk, Moscow, 108840, Russia*
[2]*Department of Physics, University of Bath, Bath BA2 7AY, United Kingdom*
[3]*Russian Quantum Center, Skolkovo, 143025, Russia*
[4]*Faculty of Physics, Lomonosov Moscow State University, Moscow, 119991, Russia*
[5]*Université Lille, CNRS, UMR 8523–PhLAM–Physique des Lasers Atomes et Molécules, F-59000 Lille, France*
[6]*ITMO University, St. Petersburg 197101, Russia*
*Corresponding author: Yaroslav.Kartashov@icfo.eu*

**We investigate theoretically frequency comb generation in a bottle microresonator accounting for the azimuthal and axial degrees of freedom. We first identify a discrete set of the axial nonlinear modes of a bottle microresonator that appear as tilted resonances bifurcating from the spectrum of linear axial modes. We then study azimuthal modulational instability of these modes and show that families of 2D soliton states localized both azimuthally and axially bifurcate from them at critical pump frequencies. Depending on detuning, 2D solitons can be either stable, or form persistent breathers, chaotic spatio-temporal patterns, or exhibit collapse-like evolution.**

Whispering gallery mode microresonators are known to generate Kerr frequency combs when sufficiently strong external pump is tuned into a resonance with one of the azimuthal modes of the resonator. Such frequency combs can be used in a number of applications, including precision spectroscopy and optical signal processing, see, e.g. [1-3]. A particularly important frequency comb state is the one, where either several or one soliton are generated as a result of phase locking of a large number of different azimuthal modes. Demonstration of soliton combs in microresonators in both anomalous [4,5] and normal [6] dispersion regimes has been a crucial step towards realization of compact solid-state sources of tunable broadband combs. Frequency combs in relatively long and thin microrings are accurately described by the one-dimensional Lugiato-Lefever (LL) model [7-10], where a single spatial dimension corresponds to the azimuthal coordinate along the resonator circumference. This model is derived and valid under an assumption that the transverse modal structure related to two remaining spatial degrees of freedom is frozen. Experimental situations, where coupling between several azimuthal families of modes corresponding to different transverse field profiles impacts comb generation and soliton formation processes in microrings are also known, see, e.g., [11]. An interplay between several spatial degrees of freedom can also be considered in other types of microresonators, such as bottle [12-15], spherical/spheroidal [16-18] and microbubble [19,20] ones. An important feature of those is that ratios between free-spectral ranges (FSRs) associated with different modal families can be more readily controlled here [15,18]. This opens new opportunities for generation of tunable frequency combs and observation of a wide spectrum of other nonlinear phenomena [15,17,18]. Considering bottle resonators two groups have recently developed a theory of frequency comb generation in them that has used 1D LL model with parabolic potential and demonstrated multistability [19] and frequency comb generation [21,22] effects relying on the axial mode family.

In this Letter we consider generation of two-dimensional frequency combs in bottle microresonators taking into account dynamics in the axial and azimuthal directions. In particular, we address a practically relevant geometry [15] where azimuthal FSR exceeds the axial one by couple of orders of magnitude, so that the corresponding resonator spectrum consists of clusters of the axial modes attached to the well separated azimuthal modes. We found 2D soliton solutions that bifurcate from the nonlinear axial modes. When these solitons destabilize through the change of the pump frequency, they transform into breathers, decay, or exhibit quasi-collapses.

We describe the dimensionless field envelope function $\Psi$ in a bottle microresonator proposing a 2D generalization of the 1D Lugiato-Lefever equation used in [21,22] to describe combs in bottle resonators. Our model takes into account an axial ($Z$ direction) trapping potential and pump localization, as well as the dependence on the azimuthal coordinate $\theta$ corresponding to the whispering gallery modes rotating around the bottle axis:

$$i\partial_T \Psi = -\frac{1}{2}d\partial_Z^2\Psi + \omega_0\Psi - i\mathcal{D}_1\partial_\theta\Psi - \frac{1}{2}\mathcal{D}_2\partial_\theta^2\Psi + \frac{f^2}{2d}\mathcal{U}(Z)\Psi - i\kappa\Psi - f\Psi|\Psi|^2 - f\mathcal{H}(Z)e^{-i\omega_p T}, \quad (1)$$

where $T$ is the physical time, $\kappa$ is the loss rate, $\omega_p$ is the pump frequency. An integrated dispersion in the azimuthal direction is approximated by $\mathcal{D}_1\mu+\mathcal{D}_2\mu^2/2$, where $\mu$ is the azimuthal mode index; $\mathcal{D}_1/2\pi$ is the azimuthal FSR; $\mathcal{D}_2$ is the second-order azimuthal dispersion. Scaling for $\Psi$, potential $\mathcal{U}$ and pump $\mathcal{H}$ is chosen in a way that the parameter $f/2\pi$ yields the axial FSR. Indeed, the linear and pump free spectrum of Eq. (1) obtained by substituting $\Psi \sim \psi_{\nu\mu}\exp(-i\omega_{\nu\mu}T-i\mu\theta)$ and assuming a parabolic trapping potential, $\mathcal{U}=Z^2$, reads as $\omega_{\nu\mu}-\omega_0=f(\nu+1/2)+\mu\mathcal{D}_1+\mu^2\mathcal{D}_2/2$, where $\nu$ is the axial index and $\psi_{\nu\mu}$ are the eigenmodes of the harmonic oscillator. Thus the spectrum is equidistant in $\nu$ and $f$ is a distance between the resonances. If a bottle microresonator has an axial length $L$, core radius $r$ and axial curvature $R$ [21,23], then the azimuthal and axial FSRs $\mathcal{D}_1 \approx c/n_0 r$, $f \approx c/n_0(rR)^{1/2}$. We set $\Psi=\psi\exp(-i\omega_p T)$ and introduce dimensionless time $t=fT$ and coordinate $z=f^{1/2}Z/d^{1/2}$ to get dimensionless equation:

$$i\partial_t\psi = -\frac{1}{2}\partial_z^2\psi + (\delta - i\beta_1\partial_\theta - \beta_2\partial_\theta^2)\psi + \frac{\mathcal{U}(z)}{2}\psi - i\gamma\psi - \psi|\psi|^2 - \mathcal{H}(z), \quad (2)$$

with detuning $\delta=(\omega_0-\omega_p)/f$, losses $\gamma=\kappa/f$, and dispersion coefficients $\beta_1=\mathcal{D}_1/f$, $\beta_2=\mathcal{D}_2/2f$. We assume that the resonator is pumped through an optical fiber running transverse to the bottle axis, which corresponds to the well localized $\mathcal{H}=h\exp[-(z-z_p)^2/w^2]$, where $w$ is the pump width ($w=0.2$ in what follows) and $z_p$ is the pump position relative to the resonator center. Accounting for spatial localization of pump is important for spectrally dense modal families, where FSR is comparable to the nonlinear shifts of the resonances, and can be disregarded for the ones with the relatively large FSRs [24]. For a microresonator with radius $r=1$ mm, curvature $R=100$ m, and length $L=2.2$ mm one obtains $\mathcal{D}_1=200$ GHz, $f=0.6$ GHz. Such resonator supports about $N=50$ axial modes. To account for the finite bottle, we set $\mathcal{U}=2N$ for $|z|>(2N)^{1/2}$. In what follows, we use $\gamma=0.004$ giving the linewidth $\kappa=2.4$ MHz, and $\beta_2=0.1$ corresponding to $\mathcal{D}_2\approx 120$ MHz. Eq. (2) was solved with the boundary conditions $\psi(z,\theta)=\psi(z,\theta+2\pi)$ and $\psi(z\to\infty,\theta)=0$.

We first consider solutions of Eq. (2) that are uniform in $\theta$, physically corresponding to the azimuthal structure of a whispering gallery mode with the frequency nearest to the pump frequency. Varying the detuning parameter we build a family of the axial modes peaking at $\delta=-(\nu+1/2)$ and shaping up a set of tilted nonlinear resonances [see Figs. 1(a),(b) showing peak amplitude $|\psi|_{max}$ of the nonlinear modes as a function of $\delta$]. When the resonator is pumped exactly at the center ($z_p=0$), then only even modes with the axial indices $\nu=0,2,...$ are excited [Fig. 1(a)]. Examples of 1D profiles of such modes can be found in [21]. Resonances are tilted and may overlap creating a multistability situation. Overlap of the central pump with the higher order harmonic oscillator modes reduces with increase of $\nu$, and hence resonance peaks gradually decrease with $\delta$, Fig. 1(a). If pump is shifted to the intensity maximum of one of the higher order modes, for example to $z_p=2.417$ (maximum of the $\nu=4$ mode), then the tilted resonances are found for odd and even modes and the modes around $\nu=4$ are excited most efficiently, see Fig. 1(b).

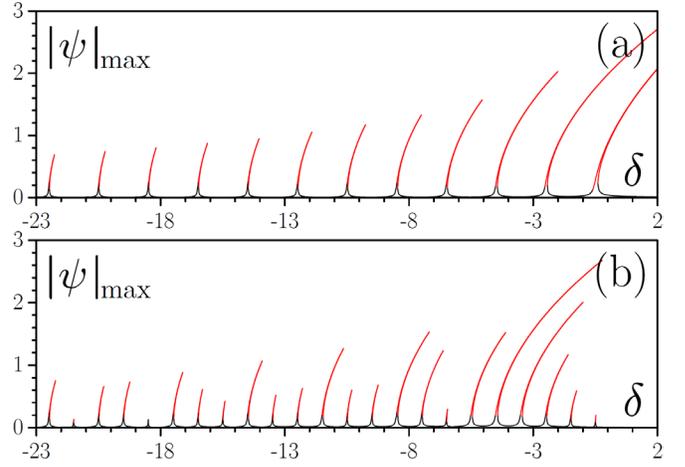

Fig. 1. A set of resonance curves corresponding to the family of nonlinear axial modes with the azimuthal index $m=0$. (a) $z_p=0$, (b) $z_p=2.417$. Here and in all figures $h=0.08$, $\gamma=0.004$.

Families of solutions shown in Fig. 1 can be subject to the instability development stimulated by small $\theta$-dependent perturbations. To analyze this effect we use an anzats $\psi(z)+u(z)e^{\lambda t+im\theta}+v^*(z)e^{\lambda^* t-im\theta}$, where $m$ is the azimuthal index and $\lambda$ is the perturbation growth rate. Linearizing Eq. (2) we find the following eigenvalue problem:

$$i\lambda u=(\delta+\beta_1 m+\beta_2 m^2+\mathcal{U}/2-i\gamma-2|\psi|^2-\partial_z^2/2)u-\psi^2 v,$$
$$i\lambda v=\psi^{*2}u-(\delta-\beta_1 m+\beta_2 m^2+\mathcal{U}/2+i\gamma-2|\psi|^2-\partial_z^2/2)v. \quad (3)$$

Solving it numerically we find a typical for the modulational instability band structure, see Fig. 2, that exists for all the upper branches of all the tilted resonances. The instability bandwidth decreases monotonically with increase of the axial mode index $\nu$ for $z_p=0$ and it varies nonmonotonically with $\nu$ for the off-center pump acquiring maximal values for strongest resonances. Unstable axial modes are indicated with the red full lines in Figs. 1 and 3. Instability of the upper branches around resonances was encountered for a broad range of $h,\gamma$ values. The instability typically breaks these modes into a complex pattern consisting of multiple solitons, sometimes with different axial structures, rotating around the bottle axis.

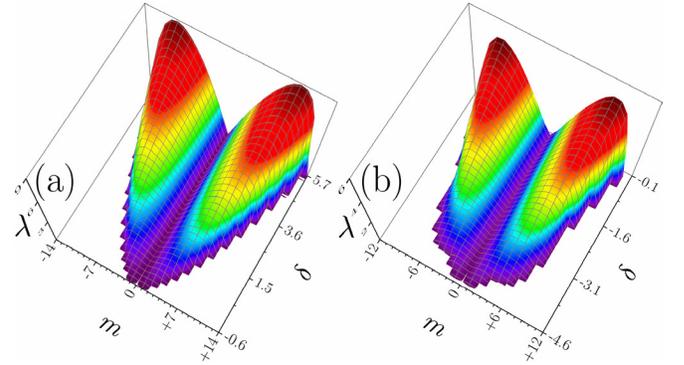

Fig. 2. Modulational instability growth rate calculated for (a) the axial mode with $\nu=0$, $z_p=0$, and for (b) $\nu=5$, $z_p=2.417$.

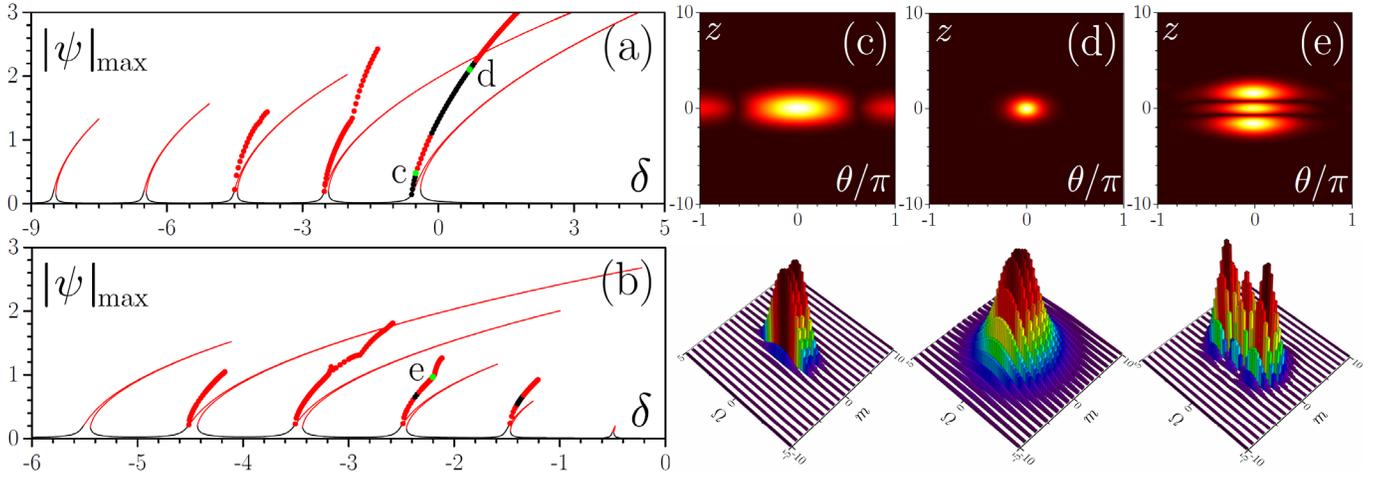

Fig. 3. (a),(b) Full lines show nonlinear resonances as in Figs. 1 (a),(b). Dots show peak amplitudes $|\psi|_{\max}$ of the 2D solitons. Stable/unstable solutions are shown black/red. (c-e) show examples of the real and frequency space profiles of the 2D solitons corresponding to points **c**, **d**, and **e** in (a) and (b).

To isolate and study the rotating 2D solitons we search for two-dimensional nonlinear modes of Eq. (2) in a moving coordinate frame, where the term $\sim \beta_1 \partial_\theta \psi$ is eliminated: $\psi(t,\theta,z) \to \psi(t,\theta',z)$ and $\theta' = \theta - \beta_1 t$. We found that families of these solitons bifurcate from the upper branches of the axial modes exactly at the detuning values, where the latter become modulationally unstable, see Figs. 3(a),(b). Azimuthal solitons we found inherit an axial structure corresponding to the respective axial modes. Close to the bifurcation point they are most delocalized in $\theta$, while they rapidly narrow down when detuning increases, see Figs. 3(c),(d). This is accompanied by substantial expansion of the spectrum in both axial ($\Omega$) and azimuthal ($m$) directions. This is particularly notable in the azimuthal direction, since the azimuthal spectrum is effectively a single mode at a bifurcation point. Examples of azimuthal spectra calculated using field distributions $\psi(\theta, z=0)$ in the center of resonator ($z=0$) are shown in Fig. 4. While solitons bifurcating from the modes with relatively small values of the axial index $\nu$ are localized very well, the ones with many axial intensity oscillations ($\nu > 2$) typically acquire complex background in the azimuthal direction with increase of $\delta$. It was unfeasible to trace all possible soliton families due to multiple bifurcations corresponding to solutions with more complex azimuthal structure.

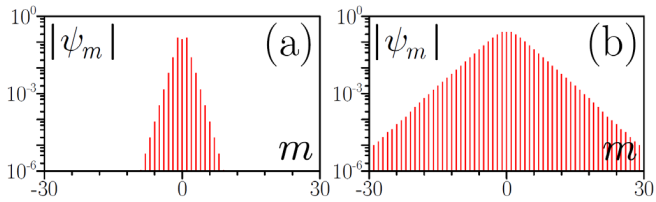

Fig. 4. Azimuthal spectra at $\delta = -0.5$ (a) and $\delta = 0.7$ (b) calculated in the point $z=0$ and corresponding to points **c**, **d** in Fig. 3(a).

However, simplest families are shown in Fig. 3. Results presented in Fig. 3 constitute the first example of 2D comb solitons. We analyzed stability of 2D comb solitons by solving Eq. (2) directly with slightly perturbed soliton inputs up to evolution times exceeding $t = 10^4$. Stable 2D soliton branches are marked black, while unstable branches are shown red. In the case of the central pump, we found a sufficiently broad interval of frequency detuning where 2D solitons connecting to the ground state of the harmonic oscillator,

$\nu = 0$, are stable. The off-centered pump only provides very narrow intervals of stability for 2D solitons connected to the $\nu = 1$ and $\nu = 2$ states of the harmonic oscillator. that are stable in narrow detuning intervals.

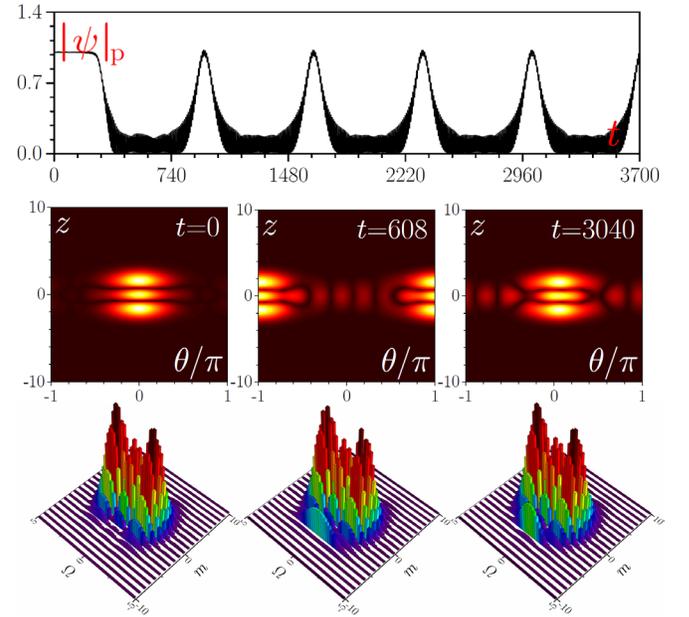

Fig. 5. 2D breather soliton around the $\nu = 2$ axial mode for $\delta = -2.1$, $z_p = 0$. Top row: field amplitude at $\theta = 0$, $z = z_p$, middle and bottom rows: real and frequency space distributions for different moments of time.

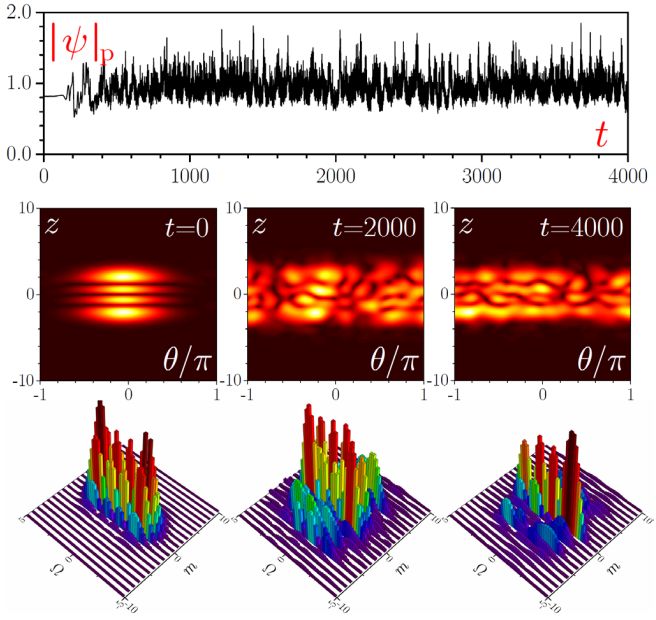

Fig. 6. Chaotic spatio-temporal dynamics emerging from the instability of the 2D soliton around the $\nu = 3$ axial mode for $\delta = -3.3$, $z_p = 2.417$. Arrangement of panels is the same as in Fig. 5.

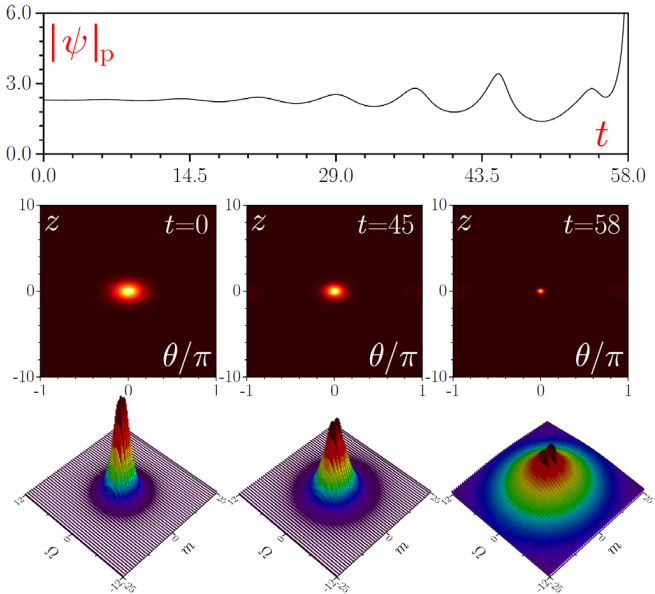

Fig. 7. Collapse of the unstable $\nu = 0$ soliton at $\delta = 0.9$, $z_p = 0$.

Unstable solitons have been found to demonstrate diverse spatio-temporal dynamics. For example, we have observed that the $\nu = 2$ solitons for $z_p = 0$ transform into persistent breathers, see Fig. 5. Note, that this transformation is accompanied by minimal modifications of the soliton spectrum and by the accumulation of the velocity off-set in the azimuthal direction. To make this offset visible we eliminated, in Fig. 5, the $\beta_1 \partial_\theta \psi$ terms corresponding to the fast soliton rotation. The small variation of FSR arising due to this offset can be estimated as $2\pi/T_b$, where $T_b$ is the time interval between subsequent peaks in Fig. 5, corresponding to arrival of moving breather to the $\theta, z_p = 0$ point. Notice that in addition to velocity offset, breather also exhibits small periodic oscillations of peak amplitude. The most common instability scenario for modes with large axial indices $\nu$ is the initial contraction followed by fragmentation of the profile into a pattern of random filaments. These chaotic patterns either persist for a very long time, see Fig. 6, or evolve into a stable nonlinear mode corresponding to the low amplitude branch of the tilted resonances. Another instability scenario of 2D solitons, which occurs when their amplitude becomes sufficiently large is the soliton collapse, which is accompanied not only by a dramatic narrowing of the field profile in both coordinates, but also by a pronounced spectral broadening in both axial and azimuthal frequencies (Fig. 7). Collapse is of course an unphysical effect and is expected to be arrested through the inclusion of the higher order dispersive terms into the model equation, which should be a subject of future research.

**Funding**: Russian Science Foundation (17-12-01413). D.V.S. acknowledges the ITMO University Visiting Professorship Scheme via the Government of Russian Federation Grant 074-U01. D.V.S., M.L.G. acknowledge EU H2020 (691011-SOLIRING) for exchange visits. D.V.S., Y.V.K. acknowledge support from the Leverhulme Trust (RPG-2015-456).